\documentclass[10pt, conference, compsocconf]{IEEEtran}
\usepackage{cite}
\usepackage{amsmath,amssymb,amsfonts}
\usepackage{graphicx}
\usepackage{textcomp}
\usepackage{xcolor}

\def\BibTeX{{\rm B\kern-.05em{\sc \kern-.025em b}\kern-.08em
    T\kern-.1667em\lower.7ex\hbox{E}\kern-.125emX}}

\usepackage{booktabs}
\usepackage{tabularx}
\usepackage{multicol}
\usepackage{multirow}
\usepackage{xspace}
\usepackage[export]{adjustbox}

\usepackage{colortbl}

\usepackage{pifont}

\usepackage{breakurl}
\usepackage[hyphens]{url}

\usepackage{algorithmic}

\usepackage{amsmath}
\usepackage{mathtools}
\usepackage[
    n,
    operators,
    advantage,
    sets,
    adversary,
    landau,
    probability,
    notions,
    logic,
    ff,
    mm,
    primitives,
    events,
    complexity,
    asymptotics,
    keys]{cryptocode}

\usepackage{fancyhdr}
\usepackage{hhline,balance}
\usepackage{datetime}
\usepackage{stfloats}
\usepackage{blkarray}

\newcommand{\squishlist}{
 \begin{list}{$\bullet$}
  { \setlength{\itemsep}{0pt}
     \setlength{\parsep}{3pt}
     \setlength{\topsep}{3pt}
     \setlength{\partopsep}{0pt}
     \setlength{\leftmargin}{1.5em}
     \setlength{\labelwidth}{1em}
     \setlength{\labelsep}{0.5em} } }

\newcommand{\squishlisttwo}{
 \begin{list}{$\bullet$}
  { \setlength{\itemsep}{0pt}
     \setlength{\parsep}{0pt}
    \setlength{\topsep}{0pt}
    \setlength{\partopsep}{0pt}
    \setlength{\leftmargin}{2em}
    \setlength{\labelwidth}{1.5em}
    \setlength{\labelsep}{0.5em} } }

\newcommand{\squishend}{
  \end{list}  }

\newcommand{\sound}{\ensuremath{\mathcal{S}}}

\newcommand{\our}{\texttt{DoubleEcho}}

\newcommand{\para}[1]{\vspace{1mm} \noindent {\bf #1}}

\PassOptionsToPackage{hyphens}{url}

\makeatletter
\def\ps@IEEEtitlepagestyle{%
  \def\@oddfoot{\mycopyrightnotice}%
  \def\@evenfoot{}%
}
\def\mycopyrightnotice{%
  {\footnotesize \textit{Proceedings of the 17th IEEE International Conference on Pervasive Computing and Communications, Kyoto, Japan, PerCom, 2019 \hfill}}
  \gdef\mycopyrightnotice{}
}

\begin{document}

\title{DoubleEcho: Mitigating Context-Manipulation Attacks in Copresence Verification}

\author{\IEEEauthorblockN{Hien Thi Thu Truong\IEEEauthorrefmark{1},
                          Juhani Toivonen\IEEEauthorrefmark{2},
                          Thien Duc Nguyen\IEEEauthorrefmark{3}, 
                          Claudio Soriente\IEEEauthorrefmark{1},
                          Sasu Tarkoma\IEEEauthorrefmark{1} and
                          N.Asokan\IEEEauthorrefmark{4}
                        }
\IEEEauthorblockA{\IEEEauthorrefmark{1}\textit{NEC Laboratories Europe}, Heidelberg, Germany}
\IEEEauthorblockA{\IEEEauthorrefmark{2}\textit{University of Helsinki}, Helsinki, Finland}
\IEEEauthorblockA{\IEEEauthorrefmark{3}\textit{Technische Universit\"at Darmstadt}, Darmstadt, Germany}
\IEEEauthorblockA{\IEEEauthorrefmark{4}\textit{Aalto University}, Espoo, Finland}
\IEEEauthorblockA{hien.truong@neclab.eu; juhani.toivonen@cs.helsinki.fi; ducthien.nguyen@trust.tu-darmstadt.de; \\ claudio.soriente@emea.nec.com; starkoma@cs.helsinki.fi; asokan@acm.org}
}

%
%
%
%
%

\IEEEoverridecommandlockouts
\IEEEpubid{\makebox[\columnwidth]{Proceedings of the 17th IEEE International Conference on Pervasive Computing and Communications, PerCom, Kyoto, Japan, 2019. 978-1-5386-5541-2/18/\$31.00~\copyright2018 IEEE \hfill} \hspace{\columnsep}\makebox[\columnwidth]{ }}

\maketitle

\IEEEpubidadjcol

\begin{abstract}
Copresence verification based on context can improve usability and strengthen security of many authentication and access control systems. By sensing and comparing their surroundings, two or more devices can tell whether they are copresent and use this information to make access control decisions. To the best of our knowledge, all context-based copresence verification mechanisms to date are susceptible to \emph{context-manipulation attacks}. In such attacks, a distributed adversary replicates the same context at the (different) locations of the victim devices, and induces them to believe that they are copresent.
In this paper we propose \our{}, a context-based copresence verification technique that leverages acoustic Room Impulse Response (RIR) to mitigate context-manipulation attacks. In \our{}, one device emits a wide-band audible chirp and all participating devices record reflections of the chirp from the surrounding environment. Since RIR is, by its very nature, dependent on the physical surroundings, it constitutes a unique \emph{location signature} that is hard for an adversary to replicate. We evaluate \our{} by collecting RIR data with various mobile devices and in a range of different locations. We show that \our{} mitigates context-manipulation attacks  whereas all other approaches to date are entirely vulnerable to such attacks. \our{} detects copresence (or lack thereof) in roughly 2 seconds and works on commodity devices.

\end{abstract}


\section{Introduction}
\label{sec:intro}


A number of authentication mechanisms leverage copresence of the (alleged) prover and the verifier to either strengthen authenticity or improve usability~\cite{Drimer:2007:KYE:1362903.1362910,Karapanos2015,DBLP:conf/esorics/HaleviMSX12,MarforioKSKC14,varshavsky07amigo}. For example, in the context of IoT, pairing of two or more devices may only be allowed among those that are copresent.
Similarly, payment systems may mitigate fraudulent in-store transactions by checking that the payment card being used and the mobile device of the legitimate card-holder are close to each other~\cite{DBLP:conf/esorics/HaleviMSX12,MarforioKSKC14}.

Copresence may be verified, for example, by exchanging an unpredictable value over a short-range communication channel. The verifier device transmits a random value over Bluetooth or NFC, and challenges the prover device to echo that same value out of band.\footnote{The out of band channel is authenticated, e.g., by means of a shared key.} However, such a na\"{i}ve solution is vulnerable to \emph{relay attacks}. The latter involves a pair of victim devices \emph{far away} from each other and a distributed attacker, copresent with each of the victims. The attacker simply relays messages between the victim devices so that they conclude to be copresent. Relay attacks have been demonstrated in research papers~\cite{francillon2010carkey,Drimer:2007:KYE:1362903.1362910} and reported in the news.\footnote{\url{http://uk.businessinsider.com/thieves-unlock-a-mercedes-using-device-relays-keys-signal-west-midlands-police-2017-11}} 

Distance bounding techniques, e.g.,~\cite{Rasmussen2010, BC93, Drimer:2007:KYE:1362903.1362910, Capkun2006} may be used against relay attacks. However, distance bounding relies on accurate measurements of round-trip times between the two devices; it therefore needs to be deployed at the lower levels of the communication stack, and often requires special hardware~\cite{Rasmussen2010}.

An alternative approach to copresence verification, that is easier to deploy than distance bounding, is based on~\emph{context}. The rationale behind context-based copresence verification is that two copresent devices should perceive similar context. Previous work has demonstrated the feasibility of using different context modalities, such as audio~\cite{DBLP:conf/esorics/HaleviMSX12, Karapanos2015, Tan2013}, radio signals~\cite{varshavsky07amigo, relay-attack-nfc}, or other features of the physical environment like humidity or pressure~\cite{ShresthaFC2014,DBLP:conf/esorics/HaleviMSX12}. For example, in~\cite{DBLP:conf/esorics/HaleviMSX12, Karapanos2015, Tan2013} two devices record environmental noise and compare their recordings to tell whether they are copresent or not.

Yet, depending on the context modality used to verify copresence, a \emph{context-manipulation attack} may be feasible. Similar to a relay attack, a context-manipulation attack involves two far away victim devices and an adversary that is copresent with each of them. Here, the adversary manipulates the context at the locations of the two devices so that they  conclude to be copresent. For example, if copresence verification is based on audio~\cite{DBLP:conf/esorics/HaleviMSX12, Karapanos2015, Tan2013}, the adversary sitting next to each of the two (far away) victim devices, may simply play an arbitrary audio clip at both locations~\cite{Shrestha:2016:SPD:2976749.2978328}; the two devices will record the same audio and conclude that they are copresent by comparing their recordings.

To the best of our knowledge, no context-based copresence verification mechanism tolerates context-manipulation attacks. For example, many copresence verification mechanisms~\cite{Narayanan11,Krumm04,varshavsky07amigo,TruongPerCom14} leverage context based on GPS, Wi-Fi and Bluetooth signals. Previous work has shown that manipulation of such context modalities is feasible~\cite{Tippenhauer2009WLanAttacks, TruongPerCom14, ShresthaTMC2018, Zeng:2017:PGL:3032970.3032983}. Audio-based copresence schemes~\cite{DBLP:conf/esorics/HaleviMSX12,Karapanos2015,Tan2013,TruongPerCom14} are also susceptible to adversarial manipulation of the context~\cite{Shrestha:2016:SPD:2976749.2978328,ShresthaTMC2018}. Copresence verification based on physical context~\cite{ShresthaFC2014,Wilson2018} is not secure to context-manipulation attacks either~\cite{ShresthaTMC2018}. Real-world applications relying on context-based copresence verification (e.g., GPS\footnote{\url{http://www.solidpass.com/authentication-methods/mobile-location-authentication.html}} or radio beacons\footnote{\url{https://securechannels.com/products/authentication/}}) are similarly vulnerable to these adversarial manipulations.

A few recent proposals~\cite{TruongPerCom14,ShresthaTMC2018} address context-manipulation adversaries, by combining multiple sensor modalities and by assuming that the adversary cannot manipulate all modalities at the same time (e.g., the adversary can manipulate Wi-Fi signals but it cannot manipulate GPS signals). Yet, if the adversary manipulates all modalities, the schemes of~\cite{TruongPerCom14,ShresthaTMC2018} become completely vulnerable to context-manipulation attacks (see Table 2 in~\cite{ShresthaTMC2018}).

\vspace{1em}
\noindent\textbf{Our contributions.} \textit{We aim at designing a context-based copresence verification mechanism that mitigates context-manipulation attacks.}
We focus on audio-based copresence verification because of the wide availability of microphones and speakers on commodity devices. In this settings, we study how to leverage the physical characteristics of the location where the protocol is executed to mitigate context-manipulation attacks. We observe that in nature many animals use echolocation to ``make sense'' of the physical environments around them. Bats emit high frequency sounds and listen to how they are reflected by the environment; dolphins also use a similar echolocation technique. Inspired by these, we make use of Room Impulse Response (RIR), which depends on both sound waves within a physical enclosure as well as \emph{the shape and the materials of the enclosure}. Sound travels via multiple paths from the source to the receiver. These paths are significantly influenced by the boundaries and the obstacles in the enclosing space. Thus, we conjecture that RIR constitutes a unique \emph{location signature} that is hard for an adversary to replicate.

We instantiate the idea above in \our{}, a mechanism for copresence verification based on RIR. \our{} mitigates context-manipulation attacks by minimizing the chances that an adversary may reproduce the same context at two different locations. In \our{}, one device emits a short (2 seconds), wide-band audible chirp and all participating devices (the playing device and one or more listening devices) record reflections of the chirp from the surrounding environment. \our{} extracts features on different frequency bands from each recording and compares those features to determine whether the devices are copresent or not. Since RIR is, by its very nature, dependent on the physical surroundings, \our{} can effectively mitigate relay and context-manipulation attacks. The same chirp signal played at two different locations yields different RIRs, unless the two locations are very similar (e.g., same shape, building materials, furniture, etc.).

We evaluate \our{} and compare it with previous work by using various mobile devices and in a range of different environments. \our{} is \textit{sound} with a false negative rate (as low as 0.021) similar to the one shown by similar proposals~\cite{Karapanos2015, Tan2013}. In face of a context-manipulation attack, \our{} is \textit{secure} with a false positive rate that ranges between 0.089 and 0.189, whereas  all other approaches to date are completely vulnerable. Finally, \our{} can be easily deployed on commodity devices --- it only requires microphones on all participating devices and a speaker on one of them --- and takes roughly 2 seconds to detect copresence.

The dataset we used for the evaluation of \our{} is publicly available for research use~\cite{rir_dataset_2018}. Source code for data collection and RIR extraction is also available upon request.

\section{Secure Contextual Copresence Verification}
\label{sec:system}


The notion of copresence may change depending on the application scenario. In some applications, copresence means that two devices are a few centimeters away (e.g., NFC payments). Others may label as copresent a pair of devices that are a few meters apart (e..g, bluetooth applications). Similar to previous work we consider that two devices are copresent if they are at most half a meter away and within the same room. This design choice suits applications such as unlocking a desktop using a mobile phone when it is in close proximity~\cite{Truong2015} and managing meeting memberships~\cite{Tan2013}.

We consider an application scenario where two devices---prover \prover{} and verifier \verifier{}--- leverage context-based copresence verification.
The protocol starts by the prover sending a copresence verification request to the verifier.
At this time both devices measure the environment via one or more available sensors and the prover sends its measurement to the verifier.
Transmission happens out-of-band on a channel authenticated, e.g., by means of a shared key. Finally, the verifier compares its measurement with the one received by the prover to decide whether the two devices are copresent. Replay protection can be achieved if the verifier sends a random nonce at the start of the protocol, and the prover piggybacks that same nonce in the measurement sent to the verifier.

In this setting, a true (resp. false) positive happens when \prover{} and \verifier{} are copresent (resp. not copresent) and \verifier's output is ``copresent''. Similarly, a true (resp. false) negative happens when \prover{} and \verifier{} are not copresent (resp. copresent) and \verifier's output is ``not copresent''.

Clearly, the copresence verification protocol should be \emph{sound}, i.e., exhibit a low rate of false negatives.

\para{Threat model.}
Let prover \prover{} and verifier \verifier{} be the two victim devices and assume they are at different locations. The goal of the adversary is to run the copresence verification protocol between \prover{} and \verifier{} and make \verifier{} conclude that the two devices are copresent.


All previous proposals for context-based copresence verification~\cite{varshavsky07amigo, Karapanos2015, DBLP:conf/esorics/HaleviMSX12, TruongPerCom14, ShresthaFC2014}  assume a standard Dolev-Yao attacker~\cite{Dolev:1981:SPK:891726}, where the adversary controls the communication network but cannot break cryptographic primitives, nor it can compromise the victim devices. Further, the adversary is not allowed to manipulate the context that the devices sense to detect copresence.

We consider a stronger adversary that is copresent with both devices and we allow for \emph{context-manipulation} attacks. That is we allow the adversary to manipulate the context at \prover{} and at \verifier. Since we use audio to verify copresence, we let the adversary control the noise in the environment of the prover and in the one of the verifier.

Therefore, the adversary succeeds if, while under attack, the copresence verification protocol outputs a \emph{false positive} (i.e., \verifier{} concludes that \prover{} is copresent while the two devices are actually far away).

Naturally, the copresence verification mechanism should be \emph{secure}, i.e., exhibit a low rate of false positives.

\section{Room Impulse Response}
\label{sec:rir}

An impulse response is the response of a system to an impulsive stimulus. In this work, we consider sound systems and their acoustic environments. By ``system'' we refer to the collective of involved parts such as the space, walls and obstacles within a room, but also the speakers and microphones on the devices.

When emitted from a speaker, a sound is distributed, albeit unevenly, in every direction, traveling multiple paths of different lengths. This results in multiple arrivals at the receiver. Sound that travels the shortest path (direct sound) arrives first and is the loudest. Sound that has traveled through other paths starts arriving soon after, with delay directly and loudness inversely proportional to the path length. Along all paths, air and surfaces absorb some of the sound energy.

An acoustic impulse response is created by how the sound is reflected along this multitude of paths; how much sound is delayed, how much sound is weakened, how much sound energy is received etc. The paths, and hence the impulse response, is heavily affected by the enclosing space such as walls, ceiling, obstacles, the power of the sound source, and the position and orientation of both the sound source and the receiver. Theoretically, two non-identical rooms should not have identical impulse responses.

\begin{figure}[t]
    \centering
    \includegraphics[scale=0.2]{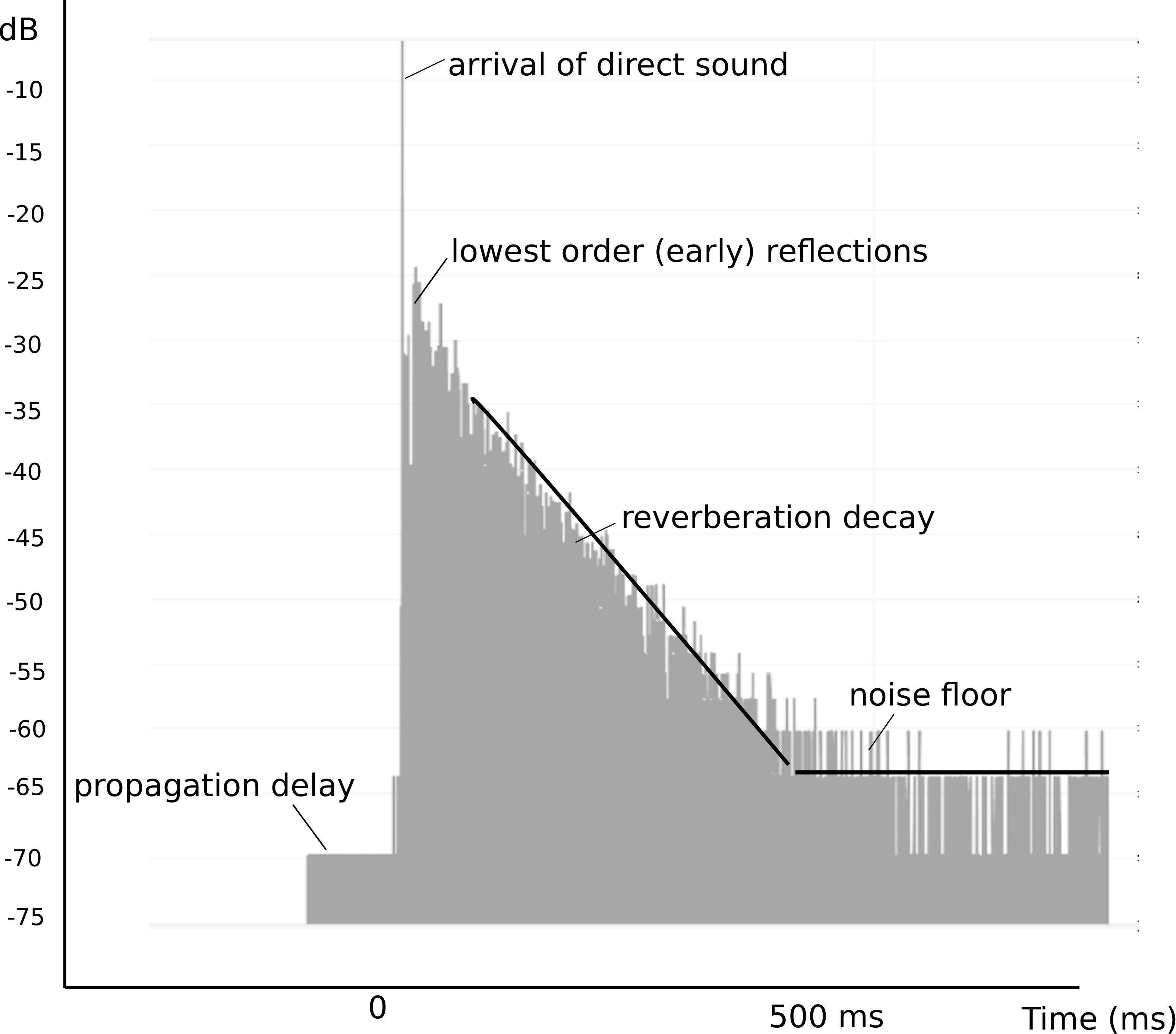}
    \caption{RIR representation in the time domain.}
    \label{fig:rir}
\end{figure}

Fig.~\ref{fig:rir} describes how an RIR looks like in the time domain. The propagation delay is the time it takes for direct sound to travel from the source to the receiver. The first part of the RIR is the arrival of the direct sound. After that, the earliest reflected sounds (lowest order reflections) start arriving, followed by sound traveling along a multitude of longer paths (higher order reflections, reverberant space). The noise floor is the point where the receiver can no longer distinguish the reverberant sound from background noise. Energy absorption by air and surface materials causes reverberant sound to decay as a function of time (and distance). In practice, the difference between direct sounds and early reflections might not be as clear as in Fig.~\ref{fig:rir}. Reverberant decay is usually measured in the range from 5 dB below the level of direct sound to a point of 60 dB below that on the backward energy curve.
\medskip

We now briefly describe standard features of RIR signals; these features are later selected by a classification model and used in \our.

Reverberation time (RT), the most common RIR feature, represents the time it takes for a reverberating signal to decay until it can no longer be distinguished from background noise. RT can be a single value even when measured for a wide band signal, for example a sweep from 20~Hz to 20~kHz. It can also be computed for different frequency bands separately.

A band covers a specific range of frequencies, with a lower bound $f_{l}$ and an upper bound $f_{h}$. Octave bands are identified by their middle frequency $f_{0}$. A band is one octave in width when the upper band frequency is twice that of the lower band frequency, \mbox{$f_{l}=f_{0}/2^{1/2}$}, \mbox{$f_{h}=f_{0}\times 2^{1/2}$}. An one-third octave band has \mbox{$f_{l}=f_{0}/(2^{1/2})^{1/3}=f_{0}/2^{1/6}$} and \mbox{$f_{h}=f_{0}\times(2^{1/2})^{1/3}=f_{0}\times 2^{1/6}$}. Reverberation time varies for different frequency bands, and hence it should be indicated if the RT applies for a specific band.

One of the most accurate method to compute RT is based on studying the decay curve. The energy curve is a curve obtained by backwards integration of the squared impulse response, which ideally starts from a point where the response falls into the noise floor.

The slope of Schroeder curve is used to measure how fast the impulse response decays. For instance, RT60 is the time it takes for a sound to decay by 60~dB. This is the standard base for measuring RT according to Sabine's reverberation equation empirically developed in the late 1890s.\footnote{\url{https://www.acoustics-engineering.com/html/sabin.html}} In practice, the level of direct sound might be less than 60~dB above the noise floor. In such cases, RT30 and RT20 are used instead. RT30 is the time a sound would decay 60~dB when extrapolated from a 30~dB decay range in the Schroeder curve (often from -5~dB to -35~dB). RT20 accordingly is extrapolated from a 20~dB decay range (from -5~dB to -25~dB).

Beside RT, there are other RIR features used for particular purposes of acoustical analysis. Early Decay Time (EDT) is derived from reverberation decay curve, conventionally in the section between 0~dB and 10~dB below the level of direct sound. EDT therefore is the reverberation time measured over the first 10~dB of the decay. It gives information of overall signal clarity and intelligibility in a room. Early-to-late energy ratio is a measure of the sound energy arriving within some specified interval after direct sound (first~$n$~milliseconds) over the energy in the remaining part of the impulse response. For example, clarity ratios C10, C35, C50, C80 are the early-to-late ratios (in dB) at 10, 35, 50, and 80 milliseconds as split times. Direct-to-Reverberant energy Ratio (DRR) is another useful measure for assessing the acoustic configuration.

\section{\our}
\label{sec:doubleecho}

\our{} has been designed for contextual copresence verification using RIR measurements. As RIR remains stable across a \emph{single} environment, it is more resilient to context-manipulation attacks than previous contextual copresence systems. In the following we describe our overall system design, after which we detail how the RIR is measured.

\subsection{System design}
\our{} uses audio as the context to verify copresence of two devices and leverages RIR to mitigate context-manipulation attacks. We focus on audio for two main reasons. First, sensors to capture audio signals (i.e., microphones) are widely available on commodity devices. Second, an audio signal is heavily affected by the surrounding space; thus, acoustic RIR may allow to tell whether two devices are within the same room.

Further, \our{} accommodates scenarios where the entropy of context data is insufficient (e.g., a silent room), by \emph{injecting} a stimulus (i.e., a chirp signal) in the environment. Therefore, we assume prover and verifier to be equipped with microphones and at least one of them to be equipped with a speaker.

The following steps show how \our{} works when run between a prover device \prover{} and a verifier device \verifier{}. We assume that \prover{} and \verifier{} have access to an authenticated channel (i.e., share a secret key).
%
\begin{enumerate}
\item \textbf{Start.} The protocol may be triggered by either \prover{} or \verifier{}. This is done by sending a well-known message over the authenticated channel. During this stage, \verifier{} also sends a random nonce to \prover{}.\footnote{The nonce is used as a challenge to prevent replay attacks.}
\item \textbf{Playing and Recording.} The device with a speaker (without loss of generality we assume \verifier{} has a speaker) plays a known audio clip \sound{}; at the same time both devices record audio in the environment via their microphones.
\item \textbf{RIR.} \verifier{} (resp. \prover{}) computes the RIR $h_{\verifier}$ (resp. $h_{\prover}$) based on the original audio clip \sound{} (the system's input) and the signal recorded via its microphone (the system's output). \prover{} sends its RIR $h_{\prover}$ to \verifier{} via the authenticated channel, along with the nonce received previously from \verifier{}.
\item \textbf{Comparison.} \verifier{} compares the nonce sent at the beginning of the protocol with the one received from \prover{}. If the two nonces are different, \verifier{} assumes a replay attack and aborts. Otherwise, \verifier{} compares $h_\verifier$ computed locally with $h_\prover$ received by \prover{} to decide whether the two devices are copresent or not.
\end{enumerate}

During the last step of the protocol, \verifier{} compares its RIR with the one received from \prover{}. The comparison mechanism clearly has a dramatic impact on the performance of the system. That is, the comparison function should enable \verifier{} to reliably detect whether \prover{} is copresent or  not. \our{} extract features from RIR and feeds them to a binary classifier that outputs a copresence verdict. 
In the remaining of this section we illustrate how we measure RIR; the next section provides details about the classifier and the features used in \our.

\subsection{Measuring RIR}
\label{sec:rir:measure}

A common method for measuring RIR is to apply a known input signal and measure the system's output.
Assuming the system to be linear time-invariant,\footnote{That is, the system exhibits a linear relationship between input and output that does not change over time.} the output signal $x(t)$ is the result of convoluting the input signal $s(t)$ and the RIR $h(t)$ (i.e., $x(t)=s(t)*h(t)$, where $t$ is time and~$*$~denotes convolution). Hence RIR can be extracted by deconvoluting input and output.

The measurement method, therefore, depends on the excitation signal and the deconvolution technique. Both the excitation signal and the deconvolution technique should maximize the Signal-to-Noise Ratio (SNR) and allow to eliminate non linear artifacts in the deconvolved impulse response.

\begin{figure}[t]
     \centering
     \includegraphics[scale=0.3]{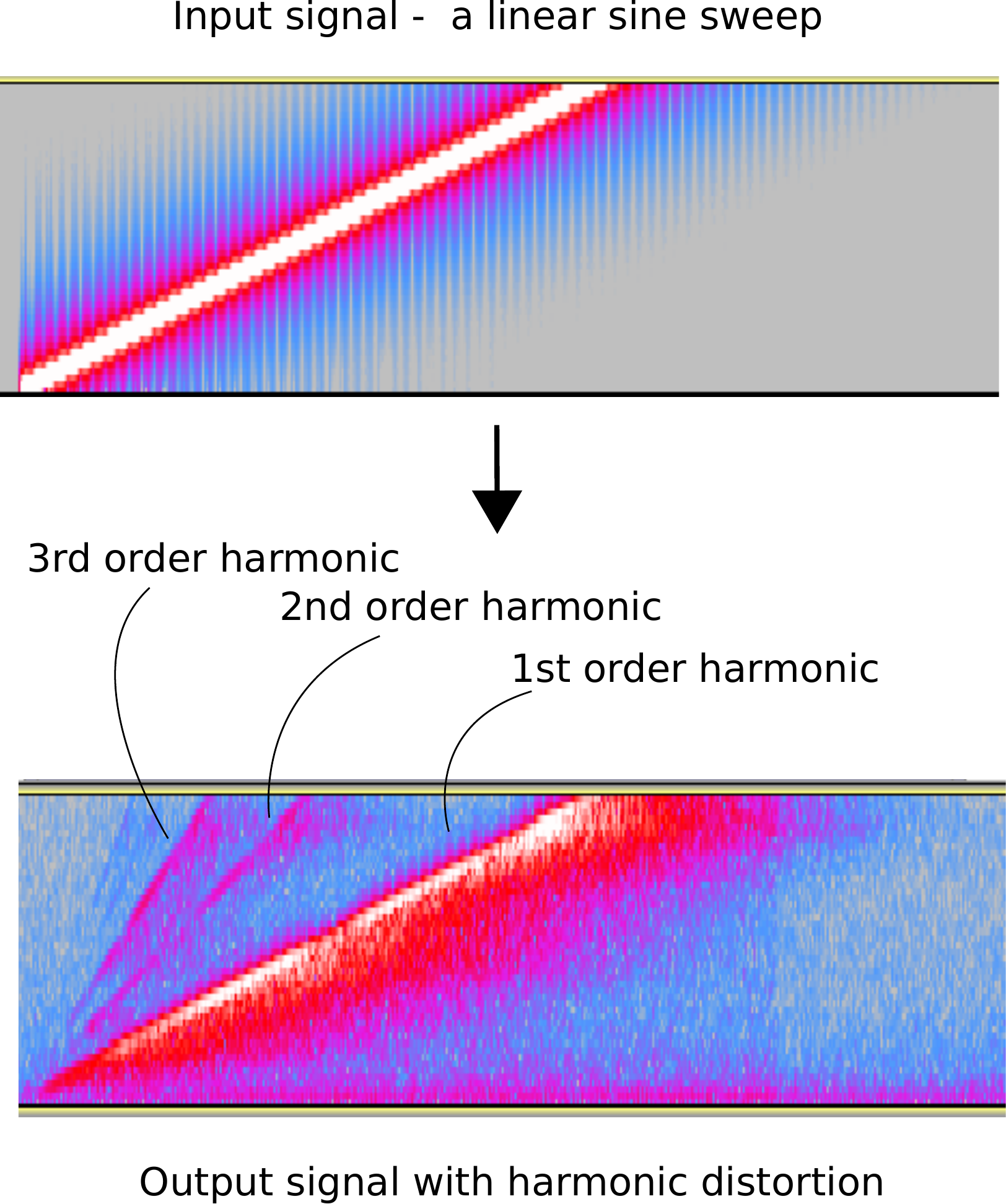}
     \caption{Distortion artifact.}
     \label{fig:harmonic}
 \end{figure}

Among various methods to measure RIR, we opt for the sine sweep technique because it overcomes the limitation of having distortion artifact when the condition of linear time variance is not fulfilled. This technique uses a linear (or an exponential) time-growing frequency sweep as the excitation signal. The output of the system in response to the sine sweep input consists of both the linear response to the excitation and the harmonic distortion at various orders due to non-linearity (Fig.~\ref{fig:harmonic}). A deconvolution of the recorded output can be used to compute the RIR. The linear impulse response is the first order harmonic in the deconvolved result.

The steps to measure RIR in \our{} are as follows: 

\begin{enumerate}
\item Convert the recorded signal (system output) $x(t)$ and the generated signal (system input) $s(t)$ from time domain to frequency domain by computing discrete Fourier transform (FFT). The FFT of a signal represents amplitudes and phases of individual frequencies in the signal.
\item Compute their deconvolution to get the transfer function. This is done by computing the time reversal of the input signal and convolve it with the recorded signal using an FFT based method~(\mbox{Matlab~fftfilt}). The transfer function shows how each frequency has been affected by the system.
\item Convert the transfer function from the frequency domain to the time domain by inverse FFT to obtain the room impulse response.

\end{enumerate}


From the RIR, we extract a series of acoustic features (RT60, EDT, D/R, C10, C35, C50, C80) across 32 frequency bands from 0 to 22050  Hz, including 1 wide-band, 10 octave bands, and 21 one-third octave bands. Features from these bands yield a vector of 224 features for each audio sample.


\section{Evaluation}
\label{sec:eval}

We started evaluation of \our{} by tuning system parameters and designing experiments. Next we ran binary classification to assess the performance of copresence verification resulted by \our{}. Given results, we conducted extra experiments to compare \our{} with previous work and elaborated its advantages.

\subsection{Parameters tuning}
\label{sec:eval:measurement}

\para{Copresence.}
The physical distance that defines copresence may vary from application to application. In our experiments, we consider that two devices are copresent  if (1) they are \textit{in the same room,} and (2) the distance between them is at most half a meter.

\para{Excitation level.} According to Section~\ref{sec:rir}, \our{} must ensure the power of the audio signal to be at least 40-50 dB above that of the noise floor level, of which 20-30 dB to measure reverberation time and 10 dB to separate reverberation and noise floor. In detail, we evaluate reverberant decay over a range, starting from 5 dB below the direct sound down to 20 or 30 dB below the starting point. One option is to emit the excitation signal at the highest level that the source device can achieve. However, if the signal is too loud for the receiver, recording will be clipped. A clipped signal is a measurement error that prevents extraction of RIR. The challenge is thus to ensure enough excitation of the acoustic space, while at the same time avoiding the clipping effect.

\para{Measurement duration.} A sweep of 500 ms or 1s is sufficient for measuring small spaces such as kitchens or offices. Larger rooms would need longer reverberation times (e.g., a couple of seconds for medium-size rooms like classrooms, up to a few seconds for large indoor spaces). In our experiments we use a sweep of two seconds to cover a wider range of rooms. In practice, if a measurement does not return good results, the parties may increase the duration of the audio clip and re-run the measurement.

\para{Audio sample and RIR features.} The signal we use is a linear sine sweep from 0 to 22050 Hz during two seconds. Recording is done at 44100 Hz 16 bit PCM. To ensure that initializing the audio system does not cut away from our signal, we pad the signal with one second of silence in the beginning and two seconds in the end, making the full sample five seconds long.

\para{Post-processing for recorded signal.} We align the recorded signal with the original generated signal (using signal cross-correlation) and remove the parts before the beginning of the sine sweep and after $500$ milliseconds past the end of the sine sweep, leaving us with just the sweep and the reverberation. We then normalize the recorded signal to account for decreased volume caused by the recording process.

\para{Linear RIR extraction.}
The first order harmonic is detected based on the highest peak in the energy decay curve. We find the highest peak and cut from 100 ms before that until 750 ms after. Background noise level is determined by energy level of 10 ms before direct sound of the first order harmonic.

\subsection{Experiment setting}
\label{sec:doubleecho:datacollection}

\para{Android application.} We developed an Android app and a backend server to facilitate collection of audio samples. The protocol is triggered by the server that instructs one device to emit a chirp signal, and all devices, including the emitter, to record. After recording, the devices send the recorded sample to the backend server.

Triggering the protocol over the network allows us to record on many devices at the same time without needing to tap buttons on their screens, and makes synchronizing the emitting and recording devices easier. To account for the different amounts of delay that the devices may experience and to make sure that the entire signal gets played through the speaker, we prepared our audio files with a few seconds of silence padding both before and after the signal. We also implemented a loudness calibration feature to avoid unwanted effects such as signal clipping.

\para{Locations and devices.} We collected data with a set of 16 smartphones and tablets spanning different brands and models (Google Nexus 5X LG, Sony Xperia Z5 compact, Sony Xperia XA, Samsung Galaxy S7, HTC One A9s, Huawei P8, ZTE, and Lenovo Phab2).
Data collection took place at two geographic locations (Helsinki and Darmstadt) and spanned 20 different rooms including offices, kitchen, corridor, classroom, meeting room, living room, bed room, stairway, computer room.

\para{Procedure.} For each room, we took a subset of the devices available, randomly chose the emitting device, and placed the remaining devices in a circle around it at a distance of roughly half a meter. We then played and recorded 5 audio samples, thereby emulating 5 executions of the copresence verification protocol.

\para{Datasets.} We split collected data in two datasets, based on whether data collection was performed in Helsinki (Dataset1) or in Darmstadt (Dataset2). For each room we used in Darmstadt, we repeated the data collection in four different locations within that room. Resulting datasets are labelled Dataset2.1, Dataset2.2, Dataset2.3, and Dataset2.4 (where Dataset2 is the union of those four). We used four locations per room in order to understand whether the location within a room affects the performance of \our.

Our data collection totalled 32975 pairs of audio samples. We discarded pairs where recording had failed for any reason (about 2\% of all data samples). Details of datasets are shown in Table.~\ref{tbl:data}; Dataset3 is the result of merging Dataset1 and Dataset2.

\begin{table}[tb]
\centering
\caption{Benign Copresent and Attack Non-Copresent Pairs.}
\begin{tabular}{lcc}
	\toprule
    Experiments & No. benign pairs & No. attack pairs \\
	\midrule
	Dataset1 & 203 & 8120 \\
    Dataset2.1 & 228 & 5725 \\
    Dataset2.2 & 189 & 3780 \\
    Dataset2.3 & 208 & 7350 \\
    Dataset2.4 & 220 & 6952 \\
    Dataset3 & 392 & 26605 \\
  \bottomrule
\end{tabular}
\label{tbl:data}
\end{table}

Similar to previous work~\cite{TruongPerCom14,ShresthaFC2014}, we use the collected audio samples to emulate benign as well as adversarial executions of the protocol. A benign execution of the protocol is one where the two devices are truly copresent. An adversarial execution happens when the two devices are not copresent.
Any pair of samples recorded by two distinct devices (where one plays an audio clip) at the same time and in the same room is considered a benign pair. Any pair of samples recorded by two distinct devices (where one also plays an audio clip) in two different rooms is considered an attack pair.

The way we build attack pairs allows us to emulate a context-manipulation attack where prover and verifier are not copresent, and the adversary plays in the prover's environment the (well known) audio clip played by the verifier. We do not emulate the random nonce sent by the verifier as a challenge. Nevertheless, an attacker copresent with both the prover and the verifier could easily relay the verifier's random nonce to the prover.

%

\subsection{Copresence classification}
\label{sec:doubleecho:classification}

\begin{table*}[htbp]
\begin{minipage}{\linewidth}
\centering
\caption{Experiment Results. Notation: Copresence \textbf{1}, Non-Copresence \textbf{0}.}

\begin{tabular}{cccccccccccccc}
	\toprule
    && \multicolumn{2}{c}{Dataset1} & \multicolumn{2}{c}{Dataset2.1} &  \multicolumn{2}{c}{Dataset2.2} &  \multicolumn{2}{c}{Dataset2.3} & \multicolumn{2}{c}{Dataset2.4} & \multicolumn{2}{c}{Dataset3} \\
    \midrule

    && \multicolumn{12}{c}{Prediction} \\
    \cmidrule{3-14}

    && {\bf 1} & {\bf 0} & {\bf 1} & {\bf 0} & {\bf 1} & {\bf 0} & {\bf 1} & {\bf 0} & {\bf 1} & {\bf 0} & {\bf 1} & {\bf 0} \\
    \midrule

    \multirow{2}{*}{Ground truth}
    &\textbf{1} & \cellcolor{gray!50} 198 & \cellcolor{gray!20} 5 & \cellcolor{gray!50} 215 & \cellcolor{gray!20} 13 &\cellcolor{gray!50} 185 &\cellcolor{gray!20} 4 &\cellcolor{gray!50} 186 &\cellcolor{gray!20} 22 &\cellcolor{gray!50} 202 &\cellcolor{gray!20} 18 &\cellcolor{gray!50} 379 &\cellcolor{gray!20} 13  \\
    &\textbf{0}&\cellcolor{gray!20} 751 &\cellcolor{gray!50} 7369 &\cellcolor{gray!20} 728 &\cellcolor{gray!50} 4997 &\cellcolor{gray!20} 336 &\cellcolor{gray!50} 3444 &\cellcolor{gray!20} 1390 &\cellcolor{gray!50} 5960 &\cellcolor{gray!20} 992 &\cellcolor{gray!50} 6708 &\cellcolor{gray!20} 2450  &\cellcolor{gray!50} 24155 \\
    \cmidrule{2-14}
    FNR  & & \multicolumn{2}{c}{0.025}  &  \multicolumn{2}{c}{0.057} & \multicolumn{2}{c}{0.021}  & \multicolumn{2}{c}{0.106} & \multicolumn{2}{c}{0.082} & \multicolumn{2}{c}{0.033} \\
    \cmidrule{2-14}
    FPR  &  &  \multicolumn{2}{c}{0.092} &  \multicolumn{2}{c}{0.127} &  \multicolumn{2}{c}{0.089} & \multicolumn{2}{c}{0.189} & \multicolumn{2}{c}{0.129} & \multicolumn{2}{c}{0.092} \\

    \bottomrule
\end{tabular}
\label{tbl:result}
\end{minipage}
\end{table*}

We consider copresence verification as a classification task for labeling two classes: \textit{copresent} (positive class) and \textit{non-copresent} (negative class). We tested different supervised learning algorithms and settled with RandomForest which performs best among all.

Our dataset is heavily imbalanced towards attack pairs and this comes from the method we use to generate our benign/attack pairs. We address the imbalance by undersampling the set of attack pairs during training; we use the RandomUnderSampler algorithm in the Scikit-learn library. We do not alter the imbalance during testing.

For each sample pair (of both benign and attack pairs), using their RIR acoustical feature vectors, we compute (component-wise) the vector of squared differences, yielding a 244-elements feature vector that are fed to the classifier. Though the feature vector has 244 elements, many of them are not informative for copresence and non-copresence classification. We applied the RandomForestRegressor algorithm implemented in Scikit-learn to select the 50 (empirically chosen among trials of 10, 20, 30, 50, 100 top features) most relevant features out of 244 available. Each training dataset we select 50-top features accordingly. Due to space limit, we do not list those selected features for each dataset.

We used 5-fold cross-validation: for each dataset, we randomly divided the data into 5 subsets (folds); 4 subsets were used for training the model and the remaining subset for testing. Table~\ref{tbl:result} shows the confusion matrix. Notably the false negative rate (FNR) ranges between 0.021 to 0.106. That is, when the two devices are copresent, \our{} will fail to detect copresence no more than 10\% of the times. Also, the false positive rate (FPR) ranges between 0.089 and 0.189 (5 over 6 datasets have FPR less than 12\%). That is, in presence of an adversary, \our{} will fail to detect attack roughly 12\% of the times.

Classification results on datasets 2.1-2.4 show that locations in rooms for copresent pairs do not affect performance of the verification mechanism. That is, when two devices are within proximity of less than half a meter, at different locations in a room, \our{} predicts copresence with similar performance.


\subsection{Comparison with previous work}
\label{sec:eval:analysis}

The effectiveness of \our{} is better appreciated when our proposal is compared to previous work.
In particular, we pick the audio-based copresence verification protocol of~\cite{TruongPerCom14} as an alternative to \our, and compare the two proposals in terms of usability, performance, and deployability. We show that \our{} provides better security without giving up usability; furthermore, requirements to deploy \our{} are the same as the ones to deploy the mechanism of~\cite{TruongPerCom14}.

Table~\ref{tbl:percom14-our-fpr} compares false positive rates (FPR) of \our{} and~\cite{TruongPerCom14} evaluated on our datasets. Recall that a false positive happens when the two devices are not copresent but the mechanism outputs a copresence verdict. From the table, it is clear that \our{} mitigates context-manipulation attacks while the mechanism of~\cite{TruongPerCom14} is completely vulnerable to such attacks. We note that the FPR reported in~\cite{TruongPerCom14} is 0.093 but the adversary in~\cite{TruongPerCom14} is not allowed context-manipulation attacks.

\looseness=-1
Table~\ref{tbl:percom14-our-fnr} compares false negative rates (FNR) of \our{} and~\cite{TruongPerCom14} evaluated on our datasets. Recall that a false negative happens when the two devices are copresent but the mechanism outputs a ``non-copresent'' verdict. A false negative hinders usability and it is desirable to have a low FNR. Both mechanisms show similar FNR across all datasets. That is, additional security of \our{} is not traded for usability.

Finally, we note that both methods have the same requirements in terms of hardware, duration, energy consumption and binary classifier.
\begin{table}[tb]
\centering
\caption{FPRs Comparison of \our{} and \cite{TruongPerCom14}.}
\begin{tabular}{lcc}
	\toprule
    Datasets & \cite{TruongPerCom14} & \our{}  \\
	\midrule
    1 & 0.995 & 0.092 \\
    2.1 & 0.996 & 0.127 \\
    2.2 & 0.999  & 0.089 \\
    2.3 & 0.995  & 0.189 \\
    2.4 & 0.998 & 0.129 \\
    3 & 0.997  & 0.092 \\
  \bottomrule
\end{tabular}
\label{tbl:percom14-our-fpr}
\end{table}

\begin{table}[tb]
\centering
\caption{FNRs Comparison of \our{} and \cite{TruongPerCom14}.}
\begin{tabular}{lcc}
	\toprule
    Datasets & \cite{TruongPerCom14} & \our{}  \\
	\midrule
    1 &  0.000 & 0.025 \\
    2.1 & 0.004 & 0.057 \\
    2.2 & 0.005  & 0.021 \\
    2.3 & 0.005  & 0.106 \\
    2.4 & 0.000 & 0.082 \\
    3 & 0.003  & 0.033 \\
  \bottomrule
\end{tabular}
\label{tbl:percom14-our-fnr}
\end{table}

We did not compare \our{} with other context-based copresence verification mechanisms that leverage audio~\cite{DBLP:conf/esorics/HaleviMSX12,Karapanos2015,Tan2013}. We could not carry out the comparison because either the code was not available, lack of detailed description prevented us from replicating the proposed mechanism, or simply the considered system model was not comparable with the one of \our.

\section{Discussion}
\label{sec:discussion}

In this section we discuss limitations of our evaluation and possible application domains of \our{}.

\looseness=-1
We have not controlled background noise levels or tested with different distances between the devices. All our experiments were made in the presence of an audible excitation signal emitted by one device. Also, devices were placed at a distance of roughly half a meter. In future work, we plan to assess how \our{} performs in presence of (loud) background noise and when devices are more than half a meter apart.

\our{} uses a wide-band audible chirp for measuring RIR. It may therefore be unsuitable for application scenarios where users my be uncomfortable with the noise \our{} generates. A possible solution may be to use ultra-sounds. Given the limited capacity of commodity devices to emit and record reflections of high frequency sounds, using ultra-sound may face several system challenges and we leave it for future work.

Our evaluation of \our{} focuses on indoor environments. We leave a feasibility study of \our{} in outdoor environments as future work. Measuring RIR in large spaces may, however, be challenging with commodity devices.

We argue that \our{} can be used in a number of application domains where it can help to mitigate context-manipulation attacks. Similar to Sound-Proof~\cite{Karapanos2015}, \our{} could be used in two-factor authentication system to enhance security while retain ease-of-use. \our{} could be also used for multiple device association~\cite{Tan2013}. During our experiments, we have empirically tested that \our{} can be effectively used to pair three or more devices at once. Finally, \our{} can be used in zero-interaction authentication systems like BlueProximity++~\cite{Truong2015} to mitigate context-manipulation attacks shown in~\cite{ShresthaTMC2018}.

Last but not least, we argue that \our{} can detect DoS attacks. With this attack, an adversary can inject audio in the environment to scramble or confuse recordings. \our{} can detect the attack by comparing the recorded signal with the priorly known input.

\section{Related Work}
\label{sec:related}

\looseness=-1
The two main directions for copresence verification in the literature are \textit{distance bounding} (radio distance bounding~\cite{Rasmussen2010, BC93, Drimer:2007:KYE:1362903.1362910, Capkun2006}; audio distance bounding~\cite{Peng2007}) and \textit{copresence verification} (RF based~\cite{varshavsky07amigo, Krumm04, Mathur11, Narayanan11}; audio~\cite{Schurmann2013, Karapanos2015, DBLP:conf/esorics/HaleviMSX12,Tan2013, shrestha2018listeningwatch}; multi-context~\cite{Miettinen2015_ConXPop, ShresthaFC2014, TruongPerCom14, Truong2015}; and other context~\cite{Czeskis08, Lester04areyou, Mayrhofer09, Castelluccia05}). In this section we discuss approaches of the latter which are related to our work. For each approach we will argue why it does not meet our design goals.

\para{Context-based copresence verification.}
Radio context such as GPS, Wi-Fi and Bluetooth is commonly used for proximity verification. Narayanan et al.~\cite{Narayanan11} studied the use of various modalities including Wi-Fi broadcast packets and access points, Bluetooth, GPS, GSM and audio atmospheric gas for private proximity detection. They concluded that Wi-Fi performs the most prominently. Krumm et al.~\cite{Krumm04} proposed ``NearMe'' which also uses Wi-Fi features for proximity detection.  Varshavsky et al.~\cite{varshavsky07amigo} presented Amigo to authenticate copresent devices using various features extracted from Wi-Fi environment. Although these solutions use the radio environment in different ways, they all depend on the availability of deployed base stations.

Acoustic copresence verification has been studied intensively. In existing solutions~\cite{Schurmann2013, Karapanos2015, DBLP:conf/esorics/HaleviMSX12, Tan2013, TruongPerCom14}, devices passively perceive ambient sound, extract its features and apply them for pairing, copresence verification and authentication. In these techniques, ambient sound is first recorded and acoustic features are then extracted.  Similarities or distances of the features, used by a classifier, are computed based on pre-defined metric. Comparing to other context such as Wi-Fi, GPS and physical environment context, audio offers several benefits. Speaker and microphone, required for producing and capturing audio, are available in most devices from computers to smartphones and wearable devices. Audio is efficient even with short recording times, only a few seconds compared to sometimes minutes for other context types such as Wi-Fi or GPS.
Halevi et al.~\cite{DBLP:conf/esorics/HaleviMSX12} proposed to use ambient audio for copresence verification of NFC devices. In their approach, ambient sounds recorded by a pair of devices in one second are compared to each other via their maximum cross correlation. 
In~\cite{Schurmann2013}, Sch\"urmann et al. proposed an approach where the devices extract an audio fingerprint as an energy matrix and compute the Hamming distance between their matrixes. Their method need at least 6 seconds of ambient audio to obtain efficient fingerprint. For automatic group membership maintenance, Tan et al.~\cite{Tan2013} assumed copresent devices form a group, and by checking similarity of silence signatures extracted from ambient sound, membership of a device can be continuously verified. 
Sound-Proof~\cite{Karapanos2015} also adopted ambient sound for two factor authentication. 

Despite benefits of using audio in copresence verification, integrity of acoustic ambience is susceptible to adversarial manipulation~\cite{Shrestha:2016:SPD:2976749.2978328}. Audio context attackers simply record audio, transmit it over a high speed channel and replay it in the victims vicinity. Furthermore, solutions in previous works largely depend on the quality of ambient sound since it is passively sensed by participating devices. In many cases where both devices are in quiet or in noisy environments, the verification result is not reliable. In addition, some approaches require to use raw audio data~\cite{DBLP:conf/esorics/HaleviMSX12, Karapanos2015}, which may raise privacy concerns. A proposed method that uses only sound signatures~\cite{Tan2013} requires long recording duration to obtain enough features. 

Physical environmental context such as temperature, humidity, gas, altitude etc. also have been studied for copresence verification. In their work~\cite{ShresthaFC2014} Shrestha et al. showed the feasibility of using such context.  

\para{RIR for localization}. 
Room level localization techniques~\cite{DBLP:journals/corr/JiaJS14, Rossi2013, Tung:2015:EAI:2789168.2790102, Wang:2014:UKS:2594368.2594384} exploit the intrinsic acoustic properties of rooms extracted from room impulse response. These approaches exploit different types of acoustic features for room localization. In RoomSense~\cite{Rossi2013} Rossi et al. showed that common audio features are superior to room acoustic features and their results were primarily based on them. SoundLoc~\cite{DBLP:journals/corr/JiaJS14} explored more acoustic features based on room impulse response including temporal features, spectral features and energetic features. EchoTag~\cite{Tung:2015:EAI:2789168.2790102} and UbiK~\cite{Wang:2014:UKS:2594368.2594384} use similar acoustic features based on characteristics of reflections of room impulse response in frequency domain to fingerprint locations. 

Even though echolocation and RIR have been used for indoor localization and room fingerprint~\cite{DBLP:journals/corr/JiaJS14, Rossi2013, Tung:2015:EAI:2789168.2790102, Wang:2014:UKS:2594368.2594384}, they have several limitations. First, whether RIR can be used for copresence verification has not been explored. Like other context-based copresence verification approaches, two devices in close proximity should perceive similar RIR. Second, previous works on room fingerprinting~\cite{DBLP:journals/corr/JiaJS14, Rossi2013} only show that the same device in different places can identify the place based on RIR, thus omitting hardware variance. For copresence verification we need to compare signals of two copresent devices. When two or more devices are involved, new challenges emerge such as hardware heterogeneity. Two different device models might behave differently even if they are in the same location. In our work, we elaborate such challenges and show that our system overcomes variance of hardware. In addition, room fingerprinting approaches require that the locations have been seen earlier. For copresence verification, this is an unreasonable assumption.

\section{Conclusion}
\label{sec:conclude}
We have presented \our{}, a copresence verification mechanism based on Room Impulse Response (RIR). To the best of our knowledge, \our{} is the only context-based copresence verification mechanism that is robust against context-manipulation attacks. In \our{}, one device emits a wide-band audible chirp and all devices record reflections of the chirp from the environment. Features extracted from the recordings are fed to a classifier to make a copresence decision.  Since RIR is, by its very nature, dependent on the physical surroundings, an adversary may not replicate the same context for two devices at different locations. By means of an Android application and an experimental campaign we have shown that RIR-based copresence verification is feasible using commodity devices. Results show that \our{} effectively mitigates context-manipulation attacks with a false positive rate as low as 0.089. While strengthening security, \our{} shows a false negative rate in line with similar proposals. \our{} operates in roughly 2 seconds and can be used with commodity devices. We expect that our proposal, experiments and evaluation results open opportunities for further applications relying on secure copresence verification.

\section*{Acknowledgment}
This work was supported by the Intel Collaborative Research Institute for Secure Computing. 
This paper has received funding from the European Union’s Horizon 2020 research and innovation programme under grant agreement No 779852.
We thank Emmanuel Vincent (INRIA) for his initial discussion and for the Matlab source code to compute RIR; Jens Matthias Bohli (NEC) and Petteri Nurmi (University of Helsinki) for their insightful discussions and comments.

\bibliographystyle{abbrv}
{\footnotesize
\bibliography{cited}
}
\end{document}